\documentclass[a4paper,11pt]{article}

\usepackage{latexsym}
\usepackage{amsmath}
\usepackage{amssymb}
\usepackage{amsthm}
\usepackage{enumerate}

\newcommand{\F}{\mathcal{F}}

\newtheorem{prop}{Proposition}[section]
\newtheorem{df}{Definition}[section]
\newtheorem{cor}{Corollary}[section]

\newcommand{\beq}{\begin{equation}}
\newcommand{\eeq}{\end{equation}}
\newcommand{\bi}{\begin{itemize}}
\newcommand{\bd}{\begin{description}}
\newcommand{\ei}{\end{itemize}}
\newcommand{\ed}{\end{description}}

\newcommand{\bc}{\begin{center}}
\newcommand{\ec}{\end{center}}

\newcommand{\sg}{\sigma}
\newcommand{\al}{\alpha}

\numberwithin{equation}{section}

\title{On multicurve models for the term structure}

\author{Laura Morino\footnote{Present affiliation: Deloitte Consulting Srl., Milano.}\\
Dipartimento di Matematica Pura ed Applicata \\
Universit\`a di Padova,  Via Trieste 63,  I-35121-Padova
\\
e-mail: laura.morino88@gmail.com\\  \\
Wolfgang J. Runggaldier\\
Dipartimento di Matematica Pura ed Applicata \\
Universit\`a di Padova,  Via Trieste 63,  I-35121-Padova\\
e-mail: runggal@math.unipd.it}

\date{}

\begin{document}

\maketitle

\begin{abstract}
In the context of multi-curve modeling we consider a two-curve setup, with one curve for discounting (OIS swap curve) and
one for generating future cash flows (LIBOR for a give tenor). Within this context we present an approach for the
clean-valuation pricing of FRAs and CAPs (linear and nonlinear derivatives) with one of the main goals being also that of
exhibiting an ``adjustment factor'' when passing from the one-curve to the two-curve setting. The model itself corresponds to short rate
modeling where the short rate and a short rate spread are driven by affine factors; this allows for correlation between short rate
and short rate spread as well as to exploit the convenient affine structure methodology. We briefly comment also on the calibration of
the model parameters, including the correlation factor.
\end{abstract}



\section{Introduction}\label{S.0}

In the wake of the big crisis one has witnessed a significant
increase in the spreads between LIBORs of different tenors as well
as the spread between a LIBOR and the discount curve (LIBOR-OIS).
This has led to the construction of {\sl multicurve models} where,
typically, future cash flows are generated through curves associated
to the underlying rates, but are discounted by another curve.

The majority of the models that have been considered reflects the
usual classical distinction between
\bd\item[i)] short rate models;
\item[ii)] HJM setup;
\item[iii)] BGM or LIBOR market models.\ed
By analogy to credit risk we may call the first two categories of
models as {\sl bottom-up} models, while the third one could be
classified as {\sl top-down}. In addition, methodologies have
appeared that are related to foreign exchange.

Here we consider only the first two setups. We begin by discussing
some issues arising with the HJM methodology and concentrate then on
short rate models. The third setup (top-down) is mainly present in
work by F. Mercurio and co-authors (see e.g. \cite{M1}, \cite{M2}),
but also in other recent work such as \cite{KRPT}. There are
advantages and disadvantages with each setup. Among the possible
advantages of short rate models is the fact that they lead more
easily to a Markovian setting, which is convenient for various
calculations (see \cite{CGNS}).  On the other hand, one of the major
advantages of HJM over a direct short rate modeling is that the
model is automatically calibrated to the initial term structure.
Short rate models in a multi-curve setup have already appeared in
the literature, e.g. \cite{KTW}, \cite{K}, \cite{FT}.

To present the basic ideas in a simple way, here we consider a {\sl
two-curve model}, namely with a curve for discounting and one for
generating future cash flows. The choice of the discount curve is
not unique; we follow the common choice of considering the {\sl OIS
swap curve}. For the risky cash flows without collateral we consider
a single LIBOR {\sl (i.e. for a given tenor structure)}.

We present an approach for the pricing of some basic LIBOR-related
derivatives, namely FRAs and CAPs ({\sl linear/nonlinear}) and
consider only {\sl clean valuation} formulas, namely without
counterparty risk. Although real pricing problems require a more
global approach (see e.g. the discussions in \cite{FST1},
\cite{FST2}, \cite{Pit}, \cite{CGGN} as well as in recent work by
D.Brigo and co-authors such as \cite{PB}, \cite{BMP}), clean
valuation formulas are nevertheless useful for various reasons: as
pointed out in \cite{CGNS}, market quotes typically reflect prices
of fully collateralized transactions so that clean price formulas
may turn out to be sufficient for calibration also when using the
model to compute possible value adjustments; furthermore (see
\cite{CGNS}), TVA adjustments are often computed on top of clean
prices. Concerning methodology, since our approach is of the
bottom-up type that considers short rate modeling, we heavily
exploit the advantages of an affine term structure. This is in
contrast with top-down approaches, where (see \cite{M1}, \cite{M2})
log-normal models are common (see however \cite{KRPT} and
\cite{GPSS} for affine LIBOR models with general distributions in a
multicurve context).

Traditionally, interest rates are defined to be coherent with the
bond prices $p(t,T)$, which represent the expectation of the market
concerning the future value of money. For the discrete compounding
forward LIBORs, which we denote here by $L(t;T,S)$, this leads to
($t<T<S$) \beq\label{1}
L(t;T,S)=\frac{1}{S-T}\,\left(\frac{p(t,T)}{p(t,S)}-1\right)\eeq
which can also be justified as representing the fair value of the
fixed rate in a FRA on the LIBOR. Since we consider only a single
LIBOR that corresponds to a given tenor structure, we assume
$S=T+\Delta$ (for tenor $\Delta$). In this way one obtains a single
curve for the term structure. The actual LIBOR rates, which in what
follows we shall denote by $\bar L(t;T,T+\Delta)$, are determined by
the LIBOR panel that takes into account various factors such as
credit risk, liquidity, etc. (see the discussion in \cite{FT}).
Following some of the recent literature, in particular \cite{CGN}
(see also \cite{KTW}), we keep the formal relationship (\ref{1})
between LIBOR rates and bond prices, but replace the risk-free bond
prices $p(t,T)$ by fictitious {\sl ``risky'' bond prices} $\bar
p(t,T)$ that are supposed to be affected by the same factors as the
actual LIBORs and that, analogously to the risk-free bond prices, we
define then  as \beq\label{3} \bar
p(t,T)=E^Q\left\{\exp\left[-\int_t^T(r_u+s_u)du\right]\mid\F_t\right\}\eeq
where $r_t$ is the classical short rate, whereas $s_t$ represents
the short rate spread ({\sl hazard rate in case of only default
risk}). Notice that in this way the spread is introduced from the
outset. Notice also that the fictitious bond prices $\bar p(t,T)$
are not actual prices.

Since in what follows we are interested in FRAs and CAPs that are
based on the $T-$spot LIBOR $\bar L(t;T,T+\Delta)$, we actually
postulate the relationship (\ref{1}) only at the inception time
$t=T$. Our starting point is thus the following relationship
\beq\label{2}\bar
L(T;T,T+\Delta)=\frac{1}{\Delta}\,\left(\frac{1}{\bar
p(T,T+\Delta)}-1\right)\eeq where we have taken into account the
fact that also for the ``risky'' bonds we have $\bar p(T,T)=1$.

In addition to the pricing of FRAs and CAPs in our two-curve setup,
our major goal here is to derive a relationship between
theoretically risk-free and actual FRAs (possibly also CAPs) thereby
exhibiting an {\sl adjustment factor} which plays a role analogous
to that of the quanto adjustments in the pricing of cross-currency
derivatives or the ``multiplicative forward basis'' in \cite{B}.

\section{The model}\label{S.1}

\subsection{Preliminary considerations}\label{S.1.1}

We start with some comments concerning HJM-like approaches to better
motivate our short rate approach. Given the bond price processes
$p(t,T)$ and $\bar p(t,T)$, in order to apply an HJM-approach, we
need to introduce corresponding forward rate processes $f^T(t)$ and
$\bar f^T(t)$ that lead to a forward rate spread expressed as
$g^T(t):=\bar f^T(t)-f^T(t)$. One then also obtains corresponding
short rates and a short rate spread, namely $r_t=f^t(t),\>\bar
r_t=\bar f^t(t),\>s_t=g^t(t)=\bar r_t-r_t$.  Notice that a
consistent model should lead to $\bar p(t,T)\le p(t,T)$, which
implies $\bar f^T(t)\ge f^T(t)$ or, equivalently $g^T(t)\ge
0\>\>\forall t<T\le\bar T,$ where $\bar T$ is a given maximal
maturity.

An extensive study within the multicurve HJM approach has appeared
in \cite{CGN}. The driving random process is a Levy and a
corresponding {\sl HJM drift condition} is derived. Conditions are
given for the non-negativity of rates and spreads; explicit formulas
are obtained for various interest rate derivatives. What may not be
fully satisfactory in \cite{CGN} is that: \bd\item[i)] some
difficulties arise when dealing not only with credit risk, but also
other risks such as liquidity. In particular, when looking for a
condition that corresponds to the {\sl defaultable HJM drift
condition};
\item[ii)] a fictitious default has to be considered explicitly
{\sl (with pre default bond prices)}.\ed
The study in \cite{CGN} is continued in the recent paper \cite{CGNS}
with the main purpose of taking into account also counterparty risk
and funding costs and of determining various {\sl valuation
adjustments} on top of the clean prices. The methodology in
\cite{CGNS} is again based on an HJM approach, but with explicit
ingredients for the induced short rate models in order to obtain a
Markovian structure and to be able to actually perform the value
adjustment calculations. In particular, the authors in \cite{CGNS}
use a Levy Hull \& White extended Vasicek model for $r_t$ and
introduce an additional factor that can be interpreted as
representing a {\sl short rate spread}. In this latter sense it
becomes analogous to the approach to be presented here.

Another HJM-based approach, limited to default risk,  appears in
\cite{CSN}  with emphasis on obtaining
Markovian models with state dependent volatilities. The driving
processes are of the jump-diffusion type. The difficulties here
appear to be given by the fact that, for convenient specifications
of the volatilities, one obtains deterministic short rate spreads.
For more general, stochastic volatilities the authors obtain only
approximate Markovianity. These difficulties have been overcome in
the subsequent paper \cite{CMS}, where the authors obtain finite-dimensional
Markovian realizations also with stochastic spreads and, in addition, obtain
a correlation structure between credit spread, interest rate and the stochastic
volatility. When trying to extend their approach to a multi curve setting, beyond
that implied by credit risk alone, there appear though some computational difficulties
due to the stochastic volatility.

Before coming now to describing our short rate model, we recall some basics concerning
FRAs. We start from the
\begin{df}\label{D.1}
A FRA ({\sl forward rate agreement}) is an OTC derivative that
allows the holder to lock in at $t<T$ the interest rate between the
inception date $T$ and the maturity $T+\Delta$ at a fixed value $K$.
At maturity $T+\Delta$, a payment based on $K$ is made and one based
on $\bar L(T;T,T+\Delta)$ is received.\end{df} We shall denote the value
of the FRA at $t<T$ by $FRA^T(t,K)$. In our two-curve risky setup,
the fair price of a FRA in $t<T$ with fixed rate $K$ and notional
$N$ is \beq\label{4} \begin{array}{lcl}
FRA^T(t,K)&=&N\Delta p(t, T+\Delta) E^{T+\Delta}\left[\bar L(T;T,T+\Delta)-K\mid \F_t\right]\\
&=&N p(t, T+\Delta)E^{T+\Delta}\left[\frac{1}{\bar{p}(T,
T+\Delta)}-(1+\Delta K)\mid \F_t\right]\end{array}\eeq where
$E^{T+\Delta}$ denotes expectation under the $(T+\Delta)-$ forward
measure $Q^{T+\Delta}$. Notice that the simultaneous presence of
$p(t, T+\Delta)$ and $\bar p(t, T+\Delta)$ does not allow for the
convenient reduction of the formula to a simpler form as in the
one-curve setup.

\subsection{Description of the model itself}\label{S.1.2}

For the {\sl short-rate model approach} we shall have to start by
modeling directly the short rate $r_t$ and the short rate spread
$s_t$ and we do it under the standard martingale measure $Q$ (to be
calibrated to the market) for the risk-free money market account as
numeraire. In order to account for a possible (negative) correlation
between $r_t$ and $s_t$ we introduce a {\sl factor model}: given
three independent affine factor processes $\Psi^i_t,\,i=1,2,3$ let
\beq\label{5}\left\{\begin{array}{lcl}
r_t&=&\Psi^2_t-\Psi^1_t\\
s_t&=&\kappa\Psi_t^1+\Psi_t^3\end{array}\right.\eeq where $\kappa$
is a constant that measures the instantaneous correlation between
$r_t$ and $s_t$ (negative correlation for $\kappa>0$). This setup
could be generalized in various ways, in particular by using more
factors to drive $s_t$. In view of the existing literature one
could, instead of using an affine model structure as we do it here,
consider e.g. ambit-type processes as presented in \cite{CFSW}. Such
a model, which is not of the semimartingale type, allows also for
analytical computations and gives the possibility to take into
account long-range dependence. Remaining within the pure credit risk
setting where, see the comment after (\ref{3}), the spread is given
by the default intensity, some of the factors affecting the spread
could be given a specific meaning as in \cite{DJ} where, using an
HJM-type approach, the authors consider a spread field process with
one of the variables representing the rating of the issuer. The
approach in \cite{DJ} could possibly be generalized also tio the
present setting.

A common approach to modeling the factors in an affine context is to
assume them of the type of a square root diffusion. This guarantees
positivity of the spread, but the negative correlation comes at the
expense of possibly negative interest rates (even if only with small
probability). With such a model, by passing to the
$(T+\Delta)-$forward measure, one can compute the value of a FRA and
of the fair fixed rate.

For various reasons, in particular in view of our main goal to
obtain an {\sl adjustment factor}, it is convenient to be able to
have the same factor model for FRAs with different maturities. We
therefore aim at  performing the calculations under a single
reference measure, namely the standard martingale measure $Q$. More
precisely, for the factor processes we assume the following affine
diffusions under $Q$ that are of the Vasicek type, namely
\beq\label{6}\left\{\begin{array}{lcl}
d\Psi_t^1&=&(a^1-b^1\Psi_t^1)dt+\sg^1\,dw_t^1\\ \\
d\Psi_t^i&=&(a^i-b^i\Psi_t^i)dt+\sg^i\sqrt{\Psi_t^i}\,dw_t^i,\quad
i=2,3\end{array}\right.\eeq where $a^i, b^i, \sg^i$ are positive
constants with $a^i\ge(\sg^i)^2/2$ for $i=2,3$, and $w_t^i$
independent Wiener processes. We have chosen a Vasicek-type model
for simplicity, but the results below can be easily extended to the
Hull \& White version of the Vasicek model. Notice that the factor
$\Psi_t ^1$ may take negative values implying that, not only $r_t$,
but also $s_t$ may become negative (see however later under
``comments on the main result''). Results completely analogous to
those that we shall obtain here for the above pure diffusion model
may be derived also for affine jump-diffusions at the sole expense
of more complicated notation.

\section{Main result (FRAs)}\label{S.2}

\subsection{Preliminary notions and results}\label{S.2.1}

Recalling the expression for a FRA under the forward measure, namely
\beq\label{7} FRA^T(t,K)=N p(t,
T+\Delta)E^{T+\Delta}\Bigg[\frac{1}{\bar{p}(T, T+\Delta)}-(1+\Delta
K)\mid \F_t\Bigg],\eeq one has that the crucial quantity to compute
is \beq\label{8}\bar\nu_{
t,T}:=E^{T+\Delta}\Bigg[\frac{1}{\bar{p}(T,
T+\Delta)}\mid\F_t\Bigg]\eeq and that the fixed rate to make the FRA
a fair contract at time $t$ is \beq\label{9} \bar K_{t}:=
\frac{1}{\Delta}(\bar\nu_{t,T }-1)\eeq In the classical single curve
case we have instead
\beq\label{10}\nu_{t,T}:=E^{T+\Delta}\Bigg[\frac{1}{{p}(T,
T+\Delta)}\mid\F_t\Bigg]=\frac{p(t,T)}{p(t,T+\Delta)}\eeq being
$\frac{p(t,T)}{p(t,T+\Delta)}$ an $\F_t-$martingale under the
$(T+\Delta)-$ forward measure. The fair fixed rate in the single
curve case is then \beq\label{11}
K_t=\frac{1}{\Delta}\left(\nu_{t,T}-1\right)
=\frac{1}{\Delta}\left(\frac{p(t,T)}{p(t,T+\Delta)}-1\right)\eeq and
notice that, in order to compute $K_t$, no interest rate model is
needed (contrary to $\bar K_t$).

Due to the {\sl affine dynamics} of $\Psi_t^i\>(i=1,2,3)$ under $Q$,
we have for the risk-free bond \beq\label{12}\begin{array}{lcl}
p(t,T)&=&E^{Q}\Big\{\exp\left[-\int_{t}^{T}r_udu\right]\mid
\F_t\Big\}=E^{Q}\Big\{\exp\left[
\int_{t}^{T}(\Psi_{u}^{1}-\Psi_{u}^{2})du\right]\mid\F_t\Big\}\\ \\
{}&=&\exp\left[A(t,T)-B^1(t,T)\Psi_t^1-B^2(t,T)\Psi_t^2\right]\end{array}\eeq
The coefficients satisfy \beq\label{13}\left\{\begin{array}{lcl}
B^1_t-b^1B^1-1=0&,&B^1(T,T)=0\\
B^2_t-b^2B^2-\frac{(\sg^2)^2}{2}(B^2)^2+1=0&,&B^2(T,T)=0\\
A_t=a^1B^1-\frac{(\sg^1)^2}{2}(B^1)^2+a^2B^2\>&,\>&A(T,T)=0\end{array}\right.\eeq
leading, in particular, to \beq\label{14}
B^1(t,T)=\frac{1}{b^1}\,\left(e^{-b^1(T-t)}-1\right).\eeq For the
{\sl risky bond} we have instead \beq\label{15}\begin{array}{lcl}
\bar
p(t,T)&=&E^{Q}\Big\{\exp\left[-\int_{t}^{T}(r_u+s_u)du\right]\mid
\F_t\Big\}\\
\\ &=&E^{Q}\Big\{\exp\left[-
\int_{t}^{T}((\kappa-1)\Psi_{u}^{1}+\Psi_{u}^{2}+\Psi_{u}^{3})du\right]\mid\F_t\Big\}\\ \\
 &=&\exp\left[\bar A(t,T)-\bar B^1(t,T)\Psi_t^1-\bar B^2(t,T)\Psi_t^2-\bar
 B^3(t,T)\Psi_t^3\right]\end{array}\eeq
This time the coefficients satisfy
\beq\label{16}\left\{\begin{array}{lcl}
\bar B^1_t-b^1\bar B^1+(\kappa-1)=0&,&\bar B^1(T,T)=0\\
\bar B^2_t-b^2\bar B^2-\frac{(\sg^2)^2}{2}(\bar B^2)^2+1=0&,&\bar B^2(T,T)=0\\
\bar B^3_t-b^3\bar B^3-\frac{(\sg^3)^2}{2}(\bar B^3)^2+1=0&,&\bar B^3(T,T)=0\\
\bar A_t=a^1\bar B^1-\frac{(\sg^1)^2}{2}(\bar B^1)^2+a^2\bar
B^2+a^3\bar B^3\>&,\>&\bar A(T,T)=0\end{array}\right.\eeq leading,
in particular, to \beq\label{17}\bar
B^1(t,T)=\frac{1-\kappa}{b^1}\,\left(e^{-b^1(T-t)}-1\right)=(1-\kappa)\,B^1(t,T)\eeq
From the above $1-$st order equations it follows that
\beq\label{18}\left\{\begin{array}{lcl}
\bar B^1(t,T)&=&(1-\kappa)\,B^1(t,T)\\ \\
\bar B^2(t,T)&=&B^2(t,T)\\ \\
\bar A(t,T)&=&A(t,T)-a^1\kappa\int_t^TB^1(u,T)du\\ \\
&{}&+\frac{(\sg^1)^2}{2}\kappa^2\int_t^T(B^1(u,T))^2du-(\sg^1)^2\kappa
\int_t^TB^1(u,T)du\\ \\
&{}&-a^3\int_t^T\bar B^3(u,T)du\end{array}\right.\eeq Letting then
\beq\label{19}\tilde A(t,T):=\bar A(t,T)-A(t,T)\eeq we obtain
\beq\label{20}\begin{array}{lcl} \bar
p(t,T)&=&\exp\Big[\bar A(t,T)-B^1(t,T)\Psi_t^1-B^2(t,T)\Psi_t^2\\
{}&{}&\hspace{5cm}-\bar B^3(t,T)\Psi_t^3+\kappa B^1(t,T)\Psi_t^1\Big]\\
{}&=&p(t,T)\,\exp\left[\tilde A(t,T)+\kappa B^1(t,T)\Psi_t^1-\bar
B^3(t,T)\Psi_t^3\right]\end{array}\eeq so that, putting for
simplicity $\tilde B^1:=B^1(T,T+\Delta)$, one may write
\beq\label{21a}\frac{p(T,T+\Delta)}{\bar
p(T,T+\Delta)}=\exp\left[-\tilde A(T,T+\Delta)-\kappa \tilde
B^1\Psi_T^1+\bar B^3(T,T+\Delta)\Psi_T^3\right].\eeq

\subsection{The result itself}\label{S.2.2}

We introduce the \begin{df}\label{D.2} We call {\sl adjustment
factor} the process
\beq\label{21}Ad_t^{T,\Delta}:=E^Q\left\{\frac{p(T,T+\Delta)}{\bar
p(T,T+\Delta)}\mid\F_t\right\},\eeq\end{df} and shall prove the
following
\begin{prop}\label{P.1}
We have \beq\label{22}\bar\nu_{t,T}=\nu_{t,T}\cdot
Ad_t^{T,\Delta}\cdot\exp\left[\kappa\frac{(\sg^1)^2}{2(b^1)^3}\left(1-e^{-b^1\Delta}\right)\left(1-e^{-b^1(T-t)}\right)^2\right]
\eeq with two {\sl adjustment factors} on the right, of which the
first one can be expressed as
\beq\label{23}\begin{array}{lcl}Ad_t^{T,\Delta}&=&e^{-\tilde
A(T,T+\Delta)}E^Q\left\{e^{-\kappa \tilde B^1\Psi_T^1+\bar
B^3(T,T+\Delta)\Psi_T^3}\mid\F_t\right\}\\&:=&A(\theta,\kappa,
\Psi_t^1, \Psi_t^3)\end{array}\eeq with
$\theta:=(a^i,b^i,\sg^i,\>i=1,2,3).$\end{prop}
One may notice the analogy here with the multiplicative forward basis in \cite{B}.

As a consequence of the previous proposition we have the following
relation between the fair value $\bar K_t$ of the fixed rate in an
actual FRA and the fair value $K_t$ in a corresponding riskless one:
\begin{cor}\label{C.1} The following relationship holds
\beq\label{24}\bar K_t=\left(K_t+\frac{1}{\Delta}\right)\cdot
Ad_t^{T,\Delta}\cdot\exp\left[\kappa\frac{(\sg^1)^2}{2(b^1)^3}\left(1-e^{-b^1\Delta}\right)\left(1-e^{-b^1(T-t)}\right)^2\right]
-\frac{1}{\Delta}\eeq\end{cor} Notice that the factor given by the
exponential is equal to $1$ for zero correlation, i.e. for
$(\kappa=0)$.

\subsection{Comments on the main result}\label{S.2.3}

\subsubsection{Comments concerning the adjustment factors}

An easy intuitive interpretation of the main result can be obtained
in the case of $\kappa=0$ (independence of $r_t$ and $s_t$): in this
case we have $r_t+s_t>r_t$ implying $\bar
p(T,T+\Delta)<p(T,T+\Delta)$ so that $Ad_t^{T,\Delta}\ge 1$ (the
exponential adjustment factor is equal to $1$). As expected, from
Proposition \ref{P.1} and Corollary \ref{C.1} it then follows that
\beq\label{25} \bar\nu_{t,T}\ge\nu_{t,T}\quad,\quad \bar K_t\ge
K_t\eeq

To gain some intuition for the cases when $\kappa\not=0$, let $\bar
p^{\kappa}(t,T), \bar\nu_{t,T}^{\kappa}, Ad_t^{T,\Delta,\kappa}$
denote the given quantities by stressing that the correlation
parameter has value $\kappa$. Notice that $p(t,T)$ and thus also
$\nu_{t,T}$ do not depend on $\kappa$. Consider then the case
$\kappa>0$, which is the standard case implying negative correlation
between $r_t$ and $s_t$. (The case $\kappa<0$ is analogous/dual).
For illustrative purposes we distinguish between the two events
$\{\Psi_t^1>0,\>\forall t\in[T,T+\Delta]\},\> \{\Psi_t^1<0,\>\forall
t\in[T,T+\Delta]\}$ where the latter occurs only with small
probability (in reality, $\Psi_t^1$ will be positive for certain
values of $t$ and negative for the remaining ones).

On $\{\Psi_t^1>0,\>t\in[T,T+\Delta]\}$ we now have
\beq\label{26}\begin{array}{l}\bar p^{\kappa}(T,T+\Delta)<\bar
p^{0}(T,T+\Delta)\\ \\
\Rightarrow\bar\nu_{t,T}^{\kappa}>\bar\nu_{t,T}^{0}\quad\Rightarrow\quad
\bar\nu_{t,T}^{\kappa}/\nu_{t,T}>\bar\nu_{t,T}^{0}/\nu_{t,T}\end{array}\eeq
Recalling then \beq\label{27} \bar\nu_{t,T}^{\kappa}=\nu_{t,T}\cdot
Ad_t^{T,\Delta,\kappa}\cdot\exp\left[\kappa\frac{(\sg^1)^2}{2(b^1)^3}\left(1-e^{-b^1\Delta}\right)\left(1-e^{-b^1(T-t)}\right)^2\right]
\eeq the last inequality in (\ref{26}) can be seen to be in line
with the fact that, in this case, in (\ref{27}) the exponential
factor is $>1$ and $Ad_t^{T,\Delta,\kappa}>Ad_t^{T,\Delta,0}$
(recall Definition \ref{D.2}).

On the other hand, on $\{\Psi_t^1<0,\>t\in[T,T+\Delta]\}$,  we have
\beq\label{28}\bar p^{\kappa}(T,T+\Delta)>\bar
p^{0}(T,T+\Delta)\quad\Rightarrow\quad
\bar\nu_{t,T}^{\kappa}/\nu_{t,T}<\bar\nu_{t,T}^{0}/\nu_{t,T}\eeq
This inequality can be seen to be in line with the fact that, here,
$Ad_t^{T,\Delta,\kappa}<Ad_t^{T,\Delta,0}$, but the exponential
factor is still $>1$. This can nevertheless be explained by noticing
that, in this case, $r_t$ is relatively large and $r_t+s_t$ is
closer to $r_t$ (may be even $<r_t$). This implies a push of
$\bar\nu_{t,T}^{\kappa}/\nu_{t,T}$ towards smaller values than in
the previous case.

\subsubsection{Comments concerning the use of the results for calibration}

For what concerns calibration of our model to FRA and other
available market data, notice that the coefficients $a^ 1,a^ 2, b^
1,b^ 2, \sg^ 1,\sg^ 2$ can be calibrated in the usual way on the
basis of the observations of default-free bonds $p(t,T)$ (if we had
a Hull\& White extension of our Vasicek-type model (\ref{6}) then
also for this model the calibration could be performed as in the
standard case). To calibrate $a^ 3,b^ 3,\sg^ 3$, notice that,
contrary to $p(t,T)$, the ``risky'' bonds $\bar p(t,T)$ are not
observable (relation (\ref{2}) does not imply a unique inverse
relationship to determine $\bar p(t,T)$ from observations of the
LIBORs). One can however observe
$K_t=\frac{1}{\Delta}\left(\frac{p(t,T)}{p(t,T+\Delta)}-1\right)$ as
well as the ``risky'' FRA rate $\bar K_t$. Recalling then Corollary
\ref{C.1} and the fact that $Ad_t^{T,\Delta}=A(\theta,\kappa,
\Psi_t^1, \Psi_t^3),$ notice that, having calibrated $a^i,b^i,\sg^i\>(i=1,2)$,
from the observations of $K_t$ and $\bar K_t$ one could thus
calibrate $a^3,b^3,\sg^3$ as well as $\kappa$. If there is a way to
determine directly  $Ad_t^{T,\Delta}$ (e.g. by observing the FRA
rates for uncorrelated $r_t$ and $s_t$), then the relationship
between $K_t$ and $\bar K_t$ as expressed in Corollary \ref{C.1}
would allow to calibrate separately $\kappa$. We furthermore recall
that, as pointed out in \cite{CGNS}, calibration of clean prices is
sufficient also when using the model to compute possible value adjustments.

\subsection{Proof of the main result}\label{S.2.4}

Since the quantities of interest, namely $\bar\nu_{t,T}$ and
$\nu_{t,T}$ were defined under the forward measure (see (\ref{8})
and (\ref{10})), as a first step we perform a change from the
forward measure $Q^{T+\Delta}$ to the standard martingale measure
$Q$. To this effect, putting $b_t:=\exp\left[\int_0^tr_udu\right]$,
the density process for changing from $Q$ to $Q^{T+\Delta}$ is
$L_t=\frac{p(t,T+\Delta)}{p(0,T+\Delta)b_t}.$ We can thus write
\beq\label{29}\begin{array}{lcl}
\bar\nu_{t,T}&=&E^{T+\Delta}\left\{\frac{1}{\bar
p(T,T+\Delta)}\mid\F_t\right\}=L_t^{-1}E^Q\left\{\frac{L_{T+\Delta}}{\bar
p(T,T+\Delta)}\mid\F_t\right\}\\ \\
&=&\frac{1}{p(t,T+\Delta)}E^Q\left\{\exp[-\int_t^{T}r_udu]\,\frac{p(T,T+\Delta)}{\bar
p(T,T+\Delta)}\mid\F_t\right\}\end{array}\eeq Recalling the
expression for $p(T,T+\Delta)/\bar p(T,T+\Delta)$ (see (\ref{21a}))
this becomes \beq\label{30}\begin{array}{l}
\bar\nu_{t,T}=\frac{1}{p(t,T+\Delta)}E^Q\Big\{e^{-\int_t^{T}r_udu}\\
\\ \quad\cdot\exp\left[-\tilde A(T,T+\Delta)-\kappa
\tilde B^1\Psi_T^1+\bar
B^3(T,T+\Delta)\Psi_T^3\right]\mid\F_t\Big\}\\
\\=\frac{1}{p(t,T+\Delta)}\exp\left[-\tilde
A(T,T+\Delta)\right]E^Q\left\{e^{\bar
B^3(T,T+\Delta)\Psi_T^3}\mid\F_t\right\}\\ \\ \hspace{3cm} \cdot
E^Q\left\{e^{-\int_t^T(-\Psi_u^1+\Psi_u^2)du}e^{-\kappa \tilde
B^1\Psi_T^1}\mid\F_t\right\}\end{array}\eeq To proceed, consider the
process $F_t$ given by the last factor in (\ref{30}), namely
\beq\label{31}
F_t:=E^Q\left\{e^{-\int_t^T(-\Psi_u^1+\Psi_u^2)du}e^{-\kappa \tilde
B^1\Psi_T^1}\mid\F_t\right\}\eeq Due to the affine dynamics of
$\Psi_t^i,\>i=1,2,$ and the independence of $\Psi_t^1$ and
$\Psi_t^2$, we may write \beq\label{32}\begin{array}{l}
F_t:=E^Q\left\{e^{\int_t^T\Psi_u^1du}e^{-\kappa
\tilde B^1\Psi_T^1}\mid\F_t\right\}E^Q\left\{e^{-\int_t^T\Psi_u^2du}\mid\F_t\right\}\\
\\
=\exp\left[\al^1(t,T)-\beta^1(t,T)\Psi_t^1\right]\,\exp\left[\al^2(t,T)-\beta^2(t,T)\Psi_t^2\right]\end{array}\eeq
where the coefficients satisfy \beq\label{33}
\left\{\begin{array}{lcl} \beta_t^1-b^1\beta^1-1=0&,&
\beta^1(T,T)=\kappa \tilde B^1\\ \\
\beta^2_t-b^2\beta^2-\frac{(\sg^2)^2}{2}(\beta^2)^2+1=0&\>,&\beta^2(T,T)=0\\
\\
\al^1_t=-\frac{(\sg^1)^2}{2}(\beta^1)^2+a^1\beta^1&,&\al^1(T,T)=0\\
\\ \al^2_t=a^2\beta^2&,&\al^2(T,T)=0\end{array}\right.\eeq
Recalling also (\ref{12})-(\ref{14}), the solutions of the system
(\ref{33}) can be expressed as \beq\label{34}\left\{
\begin{array}{lcl}\beta^1(t,T)&=&\frac{1}{b^1}\left[(b^1\kappa\tilde
B^1+1)e^{-b^1(T-t)}-1\right]=B^1(t,T)+\kappa\tilde
B^1e^{-b^1(T-t)}\\ \\
\beta^2(t,T)&=&B^2(t,T)\\ \\
\al^1(t,T)&=&\frac{(\sg^1)^2}{2}\int_t^T(\beta^1(u,T))^2du-a^1\int_t^T\beta^1(u,T)du\\
&=&\frac{(\sg^1)^2}{2}\int_t^T(B^1(u,T))^2du-a^1\int_t^TB^1(u,T)du\\
&{}&\>\>+\frac{(\sg^1)^2}{2}(\kappa \tilde B^1)^2\int_t^Te^{-2b^1(T-u)}du\\
&{}&\>\>+\kappa \tilde B^1(\sg^1)^2\int_t^TB^1(u,T)e^{-b^1(T-u)}du-a^1\kappa \tilde B^1\int_t^Te^{-b^1(T-u)}du\\ \\
\al^2(t,T)&=&-a^2\int_t^TB^2(u,T)du\end{array}\right.\eeq
Consequently \beq\label{35}
\begin{array}{l}
F_t=\exp\Bigl[\frac{(\sg^1)^2}{2}\int_t^T(B^1(u,T))^2du-a^1\int_t^TB^1(u,T)du\\
\\
\hspace{1cm}-a^2\int_t^TB^2(u,T)du-B^1(t,T)\Psi_t^1-B^2(t,T)\Psi_t^2\Big]\\
\\ \>\> \cdot\exp\Big[\frac{(\sg^1)^2}{2}(\kappa \tilde
B^1)^2\int_t^Te^{-2b^1(T-u)}du-a^1\kappa \tilde B^1\int_t^Te^{-b^1(T-u)}du\\
\hspace{6cm}-\kappa \tilde B^1e^{-b^1(T-t)}\Psi_t^1\Big]\\
\hspace{1cm}\cdot\exp\left[\kappa \tilde
B^1(\sg^1)^2\int_t^TB^1(u,T)e^{-b^1(T-u)}du\right]\\ \\
=p(t,T)\cdot\exp\Big[\frac{(\sg^1)^2}{2}(\kappa \tilde
B^1)^2\int_t^Te^{-2b^1(T-u)}du-a^1\kappa \tilde B^1\int_t^Te^{-b^1(T-u)}du\\
\hspace{6cm}-\kappa \tilde B^1e^{-b^1(T-t)}\Psi_t^1\Big]\\
\hspace{1cm}\cdot\exp\left[\kappa \tilde
B^1(\sg^1)^2\int_t^TB^1(u,T)e^{-b^1(T-u)}du\right]
\end{array}\eeq On the other hand,
recalling (\ref{21a}), one obtains \beq\label{36}\begin{array}{l}
E^Q\left\{\frac{p(T,T+\Delta)}{\bar p(T,T+\Delta)}\mid\F_t\right\}\\
\\ \quad=e^{-\tilde
A(T,T+\Delta)}E^Q\left\{e^{\bar
B^3(T,T+\Delta)\Psi_T^3}\mid\F_t\right\}E^Q\left\{e^{-\kappa \tilde
B^1\Psi_T^1}\mid\F_t\right\}\end{array}\eeq where, due to the affine
dynamics of $\Psi_t^1$, we may write
\beq\label{37}E^Q\left\{e^{-\kappa \tilde
B^1\Psi_T^1}\mid\F_t\right\}=\exp\left[\bar\al(t,T)-\bar\beta(t,T)\Psi_t^1\right]\eeq
with $\bar\al(\cdot)$ and $\bar\beta(\cdot)$ satisfying
\beq\label{38}\left\{\begin{array}{lcl}
\bar\beta_t-b^1\bar\beta=0\quad&,\quad&\bar\beta(T,T)=\kappa \tilde
B^1\\
\bar\al_t=a^1\bar\beta-\frac{(\sg^1)^2}{2}(\bar\beta)^2&,&\bar\al(T,T)=0\end{array}\right.\eeq
so that \beq\label{39}\begin{array}{lcl}
\bar\beta(t,T)&=&\kappa \tilde B^1e^{-b^1(T-t)}\\
\bar\al(t,T)&=&-a^1\kappa \tilde
B^1\int_t^Te^{-b^1(T-u)}du+\frac{(\sg^1)^2}{2}(\kappa \tilde
B^1)^2\int_t^Te^{-2b^1(T-u)}du\end{array}\eeq and,consequently,
\beq\label{40}\begin{array}{l}E^Q\left\{e^{-\kappa \tilde
B^1\Psi_T^1}\mid\F_t\right\}=\exp\left[-\kappa \tilde
B^1e^{-b^1(T-t)}\Psi_t^1\right]\\
\quad \exp\left[-a^1\kappa \tilde
B^1\int_t^Te^{-b^1(T-u)}du+\frac{(\sg^1)^2}{2}(\kappa \tilde
B^1)^2\int_t^Te^{-2b^1(T-u)}du\right]\end{array}\eeq Combining
(\ref{30}) with (\ref{35}) as well as with (\ref{36}) together with
(\ref{40}), we obtain \beq\label{41}\begin{array}{l}
\bar\nu_{t,T}=\frac{1}{p(t,T+\Delta)}\exp\left[-\tilde
A(T,T+\Delta)\right]E^Q\left\{e^{\bar
B^3(T,T+\Delta)\Psi_T^3}\mid\F_t\right\}\cdot F_t\\ \\
\qquad=\frac{p(t,T)}{p(t,T+\Delta)}\,E^Q\left\{\frac{p(T,T+\Delta)}{\bar
p(T,T+\Delta)}\mid\F_t\right\}\\
\hspace{3cm}\cdot\exp\left[\kappa(\sg^1)^2 \tilde
B^1\int_t^TB^1(u,T)e^{-b^1(T-u)}du\right].\end{array}\eeq The result
then follows noticing that \beq\label{42}\tilde
B^1\int_t^TB^1(u,T)e^{-b^1(T-u)}du=\frac{1}{2(b^1)^3}\left(1-e^{-b^1\Delta}\right)\left(1-e^{-b^1(T-t)}\right)^2.\eeq

\section{Aspects of CAP pricing}\label{S.3}

\subsection{Preliminary comments}\label{S.3.1}

This part is related to work in progress, but we want nevertheless
to present some ideas on how our results obtained for FRAs (linear
derivatives) can be extended to nonlinear derivatives. To discuss a
specific case, we concentrate here on the pricing of a {\sl single
Caplet}, with strike $K$, maturity $T$ on the spot LIBOR for the
period $[T,T+\Delta]$. Using the forward measure $Q^{T+\Delta}$, its
price in $t<T$ is then given by \beq\label{301}\begin{array}{l}
Capl^{T,\Delta}(t)=\Delta
p(t,T+\Delta)E^{T+\Delta}\left\{\left(\bar L(T;T,T+\Delta)-K\right)^+\mid\F_t\right\}\\
\\
\hspace{2cm}=p(t,T+\Delta)E^{T+\Delta}\left\{\left(\frac{1}{\bar
p(T,T+\Delta)}-\tilde K\right)^+\mid\F_t\right\}\end{array}\eeq with
$\tilde K:=1+\Delta K$.

As model, we may use the same `` risky'' short rate model as for the
FRAs that we may consider as already calibrated (for the standard
martingale measure $Q$). It may thus suffice to derive just a
pricing algorithm that need not also be used for calibration. It
remains however desirable to obtain also here an ``adjustment
factor''.

The aim, pursued in the case of the FRAs, of performing the
calculations under the same measure $Q$ leads here to some
difficulties and so we stick to forward measures.

\subsection{A possible pricing methodology}\label{S.3.2}

For the pricing, in the forward measure, we may use {\sl Fourier
transform methods} as in \cite{CGN} and \cite{CGNS} thereby
representing the claim as \beq\label{302}\left(e^X-\bar
K\right)^+\quad\mbox{with}\quad X:=-\log \bar p(T,T+\Delta)\eeq We
then need only to compute the {\sl moment generating function} of
$X$, which is a linear combination of the factors (this computation
is feasible thanks to the affine structure) and use the Fourier
transform of $f(x)=\left(e^x-\bar K\right)^+$, which is well-known.

Notice that one could possibly also apply a Gram-Charlier expansion
as in \cite{KTW}.

With the Fourier transform method the price in $t=0$ of the Caplet
can then be obtained in the form (see \cite{CGNS}) \beq\label{303}
Capl(0,T, T+\Delta)=\frac{p(0,T+\Delta)}{2\pi}\,\int \frac{\tilde
K^{1-iv-R}\bar M_X^{T+\Delta}(R+iv)}{(R+iv)\,(R+iv-1)}\,dv\eeq where
$\bar M_X^{T+\Delta}(\cdot)$ is the moment generating function of
$X$ under the $(T+\Delta)-$forward measure and $R$ is such that
$\bar M_X^{T+\Delta}(R+iv)$ is finite. This moment generating
function would have to be computed for each of the various forward
measures, but it can be directly expressed in terms of the
$Q-$characteristics of the factors: the Radon-Nikodym-derivative to
change from $Q$ to $Q^{T+\Delta}$ can in fact be expresses in
explicit form and it preserves the affine structure, see Corollary
10.2 in \cite{F} (For a recent account on conditions for an
absolutely continuous measure transformation to preserve the affine
structure see \cite{FM}).

If $M_X^{T+\Delta}(\cdot)$ is the moment generating function of $X$
with $p(T,T+\Delta)$ instead of $\bar p(T,T+\Delta)$, then
\beq\label{304}\bar M_X^{T+\Delta}(z)=M_X^{T+\Delta}(z) A(z; \theta,
\kappa, \Psi_0^1,\Psi_0^3)\eeq for a suitable $A(\cdot; \theta,
\kappa, \Psi_0^1, \Psi_0^2, \Psi_0^3)$, where $A(\cdot; \theta,
\kappa, \Psi_0^1, \Psi_0^2, \Psi_0^3)$ \linebreak
$=E^{T+\Delta}\left\{\left(M_X^{T+\Delta}\right)^{-1}\,e^{z\,X}\right\}$
which, given the affine nature of the factors, can be explicitly
computed as a function of the parameters of the model and the
initial values $\Psi_0^1, \Psi_0^2, \Psi_0^3$ of the factors. As
such, this may however not suffice to derive a satisfactory
adjustment factor as for FRAs. \vspace{.6cm}

\noindent{\bf Acknowledgements:} We are grateful to Giulio Miglietta
as well as to Claudio Fontana and Zorana Grbac for very constructive
comments.

\end{document}